\documentclass[twocolumn, prb, superscriptaddress]{revtex4}
\usepackage{color, graphicx}
\usepackage{amssymb}
\usepackage{gensymb}

\begin{document}

\title{The preparation and phase diagrams of (${^{7}}$Li${_{1-x}}$Fe${_{x}}$OD)FeSe and (Li${_{1-x}}$Fe${_{x}}$OH)FeSe superconductors}

\author{Xiuquan Zhou}
\affiliation{Department of Chemistry and Biochemistry, University of Maryland, College Park, MD 20742}
\author{Christopher K. H. Borg}
\affiliation{Department of Chemistry and Biochemistry, University of Maryland, College Park, MD 20742}
\author{Jeffrey W. Lynn}
\affiliation{NIST Center for Neutron Research, National Institute of Standards and Technology, Gaithersburg, MD 20899-6102}
\author{Shanta R. Saha}
\affiliation{Department of Physics, University of Maryland, College Park, MD 20742}
\author{Johnpierre Paglione}
\affiliation{Department of Physics, University of Maryland, College Park, MD 20742}
\author{Efrain E. Rodriguez}
\affiliation{Department of Chemistry and Biochemistry, University of Maryland, College Park, MD 20742}

\begin{abstract}
We report the phase diagram for the superconducting system (${^{7}}$Li${_{1-x}}$Fe${_{x}}$OD)FeSe and contrast it with that of (Li${_{1-x}}$Fe${_{x}}$OH)FeSe both in single crystal and powder forms. Samples were prepared via hydrothermal methods and characterized with laboratory and synchrotron X-ray diffraction, high-resolution neutron powder diffraction (NPD), and high intensity NPD. We find a correlation between the tetragonality of the unit cell parameters and the critical temperature, $T_{c}$, which is indicative of the effects of charge doping on the lattice and formation of iron vacancies in the FeSe layer. We observe no appreciable isotope effect on the maximum $T_{c}$ in substituting H by by D. The NPD measurements definitively rule out an antiferromagnetic ordering in the non-superconducting (Li${_{1-x}}$Fe${_{x}}$OD)FeSe samples below 120 K, which has been reported in non-superconducting (Li${_{1-x}}$Fe${_{x}}$OH)FeSe.$^{1}$ A likely explanation for the observed antiferromagnetic transition in (Li${_{1-x}}$Fe${_{x}}$OH)FeSe samples is the formation of impurities during their preparation such as Fe${_{3}}$O${_{4}}$ and LixFeO2, which express a charge ordering transition known as the Verwey transition near 120 K. The concentration of these oxide impurities is found to be dependent on the concentration of the lithium hydroxide reagent and the use of H${_{2}}$O vs. D${_{2}}$O as the solvent during synthesis. We also describe the reaction conditions that lead to some of our superconducting samples to exhibit ferromagnetism below $T_{c}$.
\end{abstract}

\maketitle

%%%%%%%%%%%%%%%%%%%%%%%%%%%%%%%%%%%%%%%%%%%%%%%%%%%%%%%%%%%%%%%%%%%%%%%%%%%%%%%%%%%%%%%%
%%%%%%%%%%%%%%%%%%%%%%%%%%%%%%%%%%%%%%%%%%%%%%%%%%%%%%%%%%%%%%%%%%%%%%%%%%%%%%%%%%%%%%%%
\section{Introduction}

In the iron-based pnictide and chalcogenide superconductors, chemical doping and physical pressure are universal variables by which to tune the superconducting properties.$^{2, 3}$ For example, the critical temperature, $T_{c}$, of 8 K in FeSe under ambient conditions$^{4-7}$ can be raised to 38 K by externally applied pressure$^{8, 9}$ or 44 K by intercalation of cationic species.$^{10-13}$ The tetragonal ($P4/nmm$) structure of FeSe (Figure 1) consists of sheets of edge-sharing FeSe$_{4}$ tetrahedra held together by van der Waals interactions, which makes it an ideal host for intercalation chemistry. Negative pressure, or strain, has also been implicated as a parameter in the high $T_{c}$ of 65 K - 100 K reported for single layered FeSe.$^{14-16}$
Given the propensity of the FeSe layered system for chemical and physical manipulation, FeSe is an ideal platform for under-standing the superconductivity of the iron-based systems and for the preparation of new layered functional materials. The recently discovered (Li${_{1-x}}$Fe${_{x}}$OH)FeSe system,$^{1, 17-21}$ which contains PbO-type layers of LiOH alternating with the anti-PbO type layers of FeSe (Figure 1), offers such an opportunity. Iron may occupy the lithium site and therefore effectively charge dope the FeSe layer since the (Li${_{1-x}}$Fe${_{x}}$OH) layer would be positively charged.  Sun \textit{{\it et al}}. have also reported that increased lithiation of the (Li${_{1-x}}$Fe${_{x}}$OH) layer would force iron to occupy any vacancies in the FeSe layer, which can be detrimental to the superconducting properties.$^{18}$

\begin{figure}[b]
\centering
  \includegraphics[width=1\columnwidth]{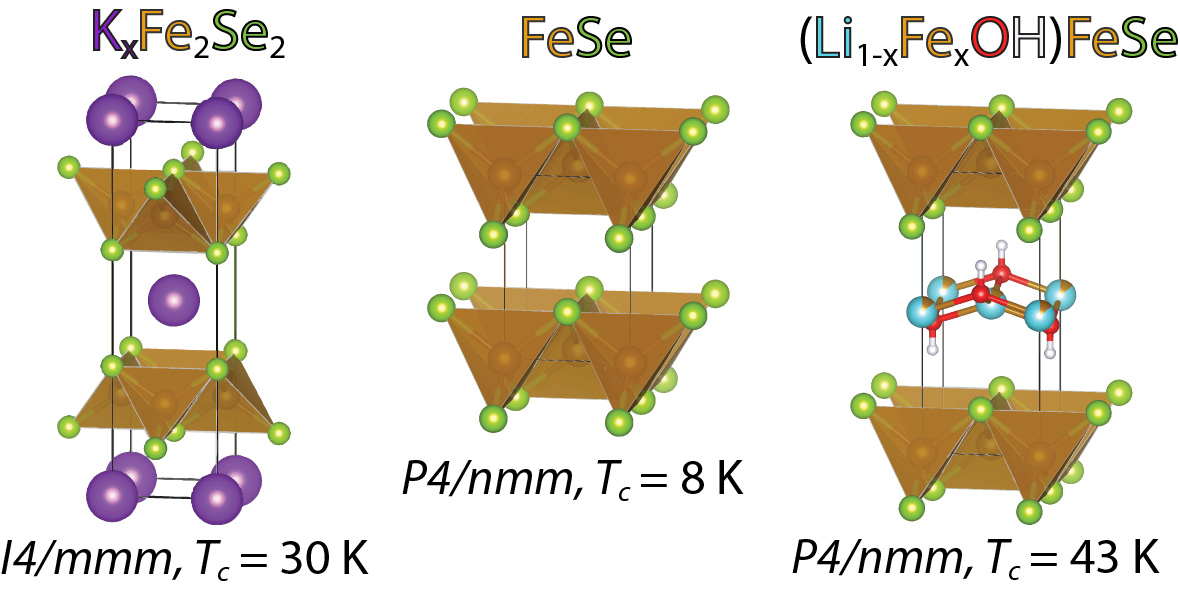}
  \caption{Crystal structures of three layered iron selenides: K${_{0.8}}$Fe${_{1.6}}$Se${_2}$ (left) FeSe (centre) and (Li${_{1-x}}$Fe${_{x}}$OH)FeSe (right)}
  \label{fgr_1}
\end{figure}

Three outstanding issues in the (Li${_{1-x}}$Fe${_{x}}$OH)FeSe system and related phases are whether 1) the parent phase is antiferromagnetic, 2) superconductivity coexists with ferromagnetism, and 3) any isotope effects on $T_{c}$ exist. Critical to answering all three questions is the preparation of the deuteroxide version of (Li${_{1-x}}$Fe${_{x}}$OH)FeSe and comparing their phase diagrams. Furthermore, hydrothermal synthesis under either H${_{2}}$O or D${_{2}}$O presents interesting differences in the purity of the resulting superconducting phases due to differences in the reaction kinetics. Thus, this study will help workers in the field understand the thermodynamic and kinetic factors in the preparation of phase pure and superconducting FeSe-based materials.

Due to the large incoherent scattering of hydrogen and high neutron absorption cross section of ${^{6}}$Li, our compounds were prepared doubly isotopically pure, (${^{7}}$Li${_{1-x}}$Fe${_{x}}$OD)FeSe, which allowed for an opportunity to complete a phase diagram for the deuterated series. Although some studies have found no evidence of ferromagnetism below $T_{c}$ in their samples,.$^{1, 20}$ under the right synthesis conditions, we have observed a ferromagnetic signal in the superconducting regime as first reported by Pachmayr \textit{{\it et al}}.$^{17}$ Herein, we report the phase diagram for (Li${_{1-x}}$Fe${_{x}}$OD)FeSe, compare it to that of (Li${_{1-x}}$Fe${_{x}}$OH)FeSe, and investigate the magnetic properties of the non-superconducting and superconducting samples.

%%%%%%%%%%%%%%%%%%%%%%%%%%%%%%%%%%%%%%%%%%%%%%%%%%%%%%%%%%%%%%%%%%%%%%%%%%%%%%%%%%%%%%%%
%%%%%%%%%%%%%%%%%%%%%%%%%%%%%%%%%%%%%%%%%%%%%%%%%%%%%%%%%%%%%%%%%%%%%%%%%%%%%%%%%%%%%%%%
\section{Experimental}

\subsection{Sample preparation}
The preparation of powder samples were modified from a hydrothermal route reported in the literature.$^{1, 22}$ For the synthesis of deuterated samples we first prepared the doubly isotopically pure $^{7}$LiOD as a precursor. $^{7}$LiOD was prepared by mixing a stoichiometric amount of $^{7}$LiCO$_{3}$ (Sigma Aldrich, 99\% for $^{7}$Li) and CaO (calcined from CaCO$_{3}$, Sigma Aldrich, 99\%) in D${_{2}}$O. The CaCO$_{3}$ precipitate was filtered, and $^{7}$LiOD crystallized by evaporation of the solvent.

For a typical preparation of (Li${_{1-x}}$Fe${_{x}}$OD)FeSe, 5 mmol of Fe powder (Alfa Aesar, 99.9\%), 6 mmol of selenourea (Sigma Aldrich, 98\%) and 50 mmol of LiOD powder were suspended in 5 mL of distilled D${_{2}}$O (Oxford Isotope, 99.9\%). The mixture was placed in a Teflon-lined stainless steel autoclave at 120-200 $^\circ$C for 3-5 days. Afterwards, the autoclave was opened in an argon-filled glove bag, and the shiny black precipitate was washed with D${_{2}}$O. The product was washed and centrifuged several times until the supernatant was clear. The remaining product was collected, vacuumed dried, and stored in a nitrogen-filled glove box.  The yield of the product was usually between 50\% and 70\%.

Single crystal (Li${_{1-x}}$Fe${_{x}}$OD)FeSe and (Li${_{1-x}}$Fe${_{x}}$OH)FeSe samples were prepared by replacing potassium cations with LiOD or LiOH from K${_x}$Fe${_{2-y}}$Se${_2}$ single crystals under hydrothermal conditions similar to those reported by Dong \textit{{\it et al}}.$^{23}$ For the growth of the K${_x}$Fe${_{2-y}}$Se${_2}$ single crystals, 1.8 g (13 mmol) of FeSe powder was mixed with 0.21 g (5.4 mmol) of potassium metal (Alfa Aesar, 99\%) to match the nominal composition of K${_{0.8}}$Fe${_{2-y}}$Se${_2}$.$^{24, 25}$ The FeSe precursor was prepared by heating Fe (Alfa Aesar, 99.9\%) and Se (Alfa Aesar, 99\%) powders to 700 $^\circ$C for 5 h followed by furnace cooling; the resulting phase does not need to be of the tetragonal $\beta$-FeSe form for the crystal growth. The FeSe/K mixtures were loaded in a quartz ampoule inside a nitrogen-filled glovebox, and the ampoules flame sealed under vacuum. In order to avoid oxidation of the samples from breaking of the ampoule due to potassium-induced corrosion of the quartz walls, the sample container was sealed in a larger ampoule. For crystal growth of K${_x}$Fe${_{2-y}}$Se${_2}$, the mixture was heated to 1030 $^\circ$C over 10 h and held at 1030 $^\circ$C for 3 hours to form a homogeneous melt. Subsequently, the melt was slowly cooled at a rate of 6 $^\circ$C/hour to 650 $^\circ$C to allow crystal growth. After cooling to room temperature, K${_x}$Fe${_{2-y}}$Se${_2}$ single crystal approximately 8 mm in diameter was recovered.

In order to compare the effect of D${_{2}}$O to the reaction kinetics, single crystals of both (Li${_{1-x}}$Fe${_{x}}$OH)FeSe and (Li${_{1-x}}$Fe${_{x}}$OD)FeSe were prepared under identical hydrothermal conditions. For the preparation of (Li${_{1-x}}$Fe${_{x}}$OH)FeSe single crystals, the K${_x}$Fe${_{2-y}}$Se${_2}$ precursor (0.2 g - 0.4 g), 0.14 g (2.5 mmol) Fe powder, and 2 g (47 mmol) LiOH monohydrate were added to 5 mL water. For (Li${_{1-x}}$Fe${_{x}}$OD)FeSe single crystals, to match the concentration of LiOH in water, 1.2 g (47 mmol) LiOD and 6 mL D${_{2}}$O were used for reactions. The mixture was placed in a Teflon-lined stainless steel autoclave at 120-200 $^\circ$C for 4-5 days. Silver colored single crystals were recovered by washing away excess powder with water and drying under vacuum overnight. The as-recovered single crystals retained similar shape to the starting K${_x}$Fe${_{2-y}}$Se${_2}$ single crystals. 

Samples prepared in the absence of excess iron powder were not superconducting, which could be due to either oxidation of the iron or vacancy formation in the FeSe layer. To study the role of metal powders during the cation exchange reactions, experiments using Sn metal (Alfa Aesar, 99.9\%) instead of Fe powder for the preparation of ((Li${_{1-x}}$Fe${_{x}}$OH)FeSe and ((Li${_{1-x}}$Fe${_{x}}$OD)FeSe single crystals at 120 $^\circ$C were carried out. Sn can react with hot concentrated bases to form soluble [Sn(OH)$_{6}$]$^{2+}$ species while evolving H${_{2}}$ gas,${^{26}}$ thus providing a stronger reducing environment than the hydrothermal reactions without the presence of metal powders.

\subsection{Laboratory and synchrotron X-ray diffraction measurements}

%%%%%%%%%%%%%%%%%%%%%%%%          Figure 2   %%%%%%%%%%%%%%%%%%%%%%%%%%%%%%%%%%%%%%%%%%
\begin{figure}[b]
\centering
  \includegraphics[width=1\columnwidth]{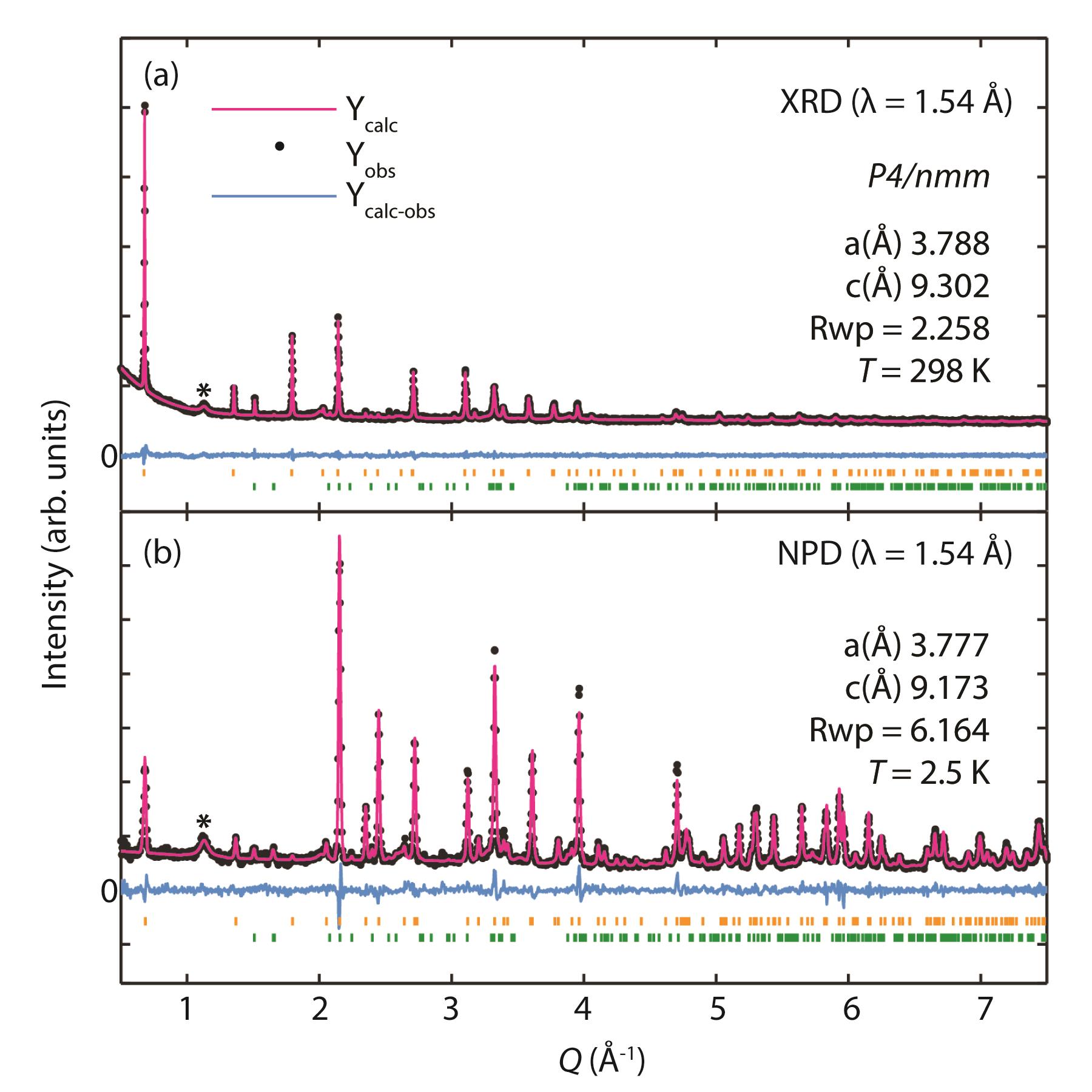}
  \caption{(a) X-ray and (b) neutron powder diffraction for non-superconducting (${^{7}}$Li${_{1-x}}$Fe${_{x}}$OD)FeSe with x = 0.166. No magnetic phase could be indexed in the NPD, indicating lack of antiferromagnetic ordering. Weight percent fractions from structural refinements are as follows: 98\% (${^{7}}$Li${_{1-x}}$Fe${_{x}}$OD)FeSe (gold ticks) and 2\% Li${_{2}}$CO${_{3}}$ (green ticks). A few broad peaks corresponding to FeSe were fit by a Pawley routine (asterisk).}
  \label{fgr_2}
\end{figure}

Room temperature powder X-ray diffraction (PXRD) data were collected on a Bruker D8 X-ray diffractometer (Cu Kα radiation, $\lambda$ = 1.5418 \AA). Data were collected with a step size of 0.02$^\circ$ between 7$^\circ$ and 70$^\circ$ for Pawley fits to extract lattice constants and 7$^\circ$ and 120$^\circ$ for Rietveld fits to obtain better structural parameters. 
In order to find any possible crystallographic phase transitions that are coupled to either the superconducting or magnetization order parameters, temperature dependent (5-300 K) high-resolution synchrotron X-ray diffraction was carried out for powders of ground single crystals at Beamline

11-BM at the Advanced Photon Source (APS). In addition to finding subtle changes in symmetry due to peak splitting, the synchrotron measurements provide high-$Q$ data and therefore more accurate structural parameters. Analysis of the high-$Q$ reflections help determine any small changes in the iron occupancies in both the FeSe and LiOH layers, which could affect $T_{c}$'s of the sample as suggested by Sun  \textit{{\it et al}}.$^{18}$ An Oxford helium cryostat (closed flow system) was used to reach a temperature that is close to liquid helium ($\approx$ 4 K). Ground powders of single crystals were packed in 0.4 mm Kapton capillaries tubes and sealed with epoxy. Diffraction data were collected between 0.5$^\circ$ and 46$^\circ$ with a step size of 0.0001$^\circ$ using a constant wavelength λ = 0.413964 \AA \ (30 keV). 

\subsection{Magnetization measurements}
Magnetic susceptibility measurements were carried out using a magnetic property measurement system (Quantum Design 
MPMS). Both field-cooled (FC) and zero-field-cooled (ZFC) magnetic susceptibility measurements were taken from 2 K - 300 K in direct current mode with an applied magnetic field of 1 or 3 mT. 

\subsection{Neutron powder diffraction measurements}

All the neutron work was carried out with doubly isotopically pure samples (${^{7}}$Li${_{1-x}}$Fe$_{x}$OD)FeSe at the NIST Center for Neutron Research (NCNR). The samples were loaded into He-filled vanadium cans and subsequently into a closed cycle refrigerator for low temperature measurements. Low temperature (4 K) diffraction data were collected on the BT-1 high-resolution NPD with the Cu(311) monochromator ($\lambda$ = 1.540 \AA). In addition to base temperature measurements, we performed NPD measurements at various temperatures 25 K, 75 K, 150 K, and room temperature to search for any crystallographic phase transitions.  
High-intensity and coarse-resolution diffraction measurements were carried out on the BT-7 spectrometer ($\lambda$ = 2.359 \AA) using the position sensitive detector (PSD) to search for magnetic Bragg peaks from base temperature up to 150 K.$^{27}$

%%%%%%%%%%%%%%%%%%%%%%%%%%%%%%%%%%%%%%%%%%%%%%%%%%%%%%%%%%%%%%%%%%%%%%%%%%%%%%%%%%%%%%%%
%%%%%%%%%%%%%%%%%%%%%%%%%%%%%%%%%%%%%%%%%%%%%%%%%%%%%%%%%%%%%%%%%%%%%%%%%%%%%%%%%%%%%%%%
\section{Results}

%%%%%%%%%%%%%%%%%%%%%%%%%%%%%%%%%%%%%%%%%%%%%%%%
\subsection{Crystallography and phase diagram}

Rietveld refinements with both XRD and NPD data were carried out with the TOPAS 4.2 software.$^{28}$  Representative fits to one of the deuteroxide samples are presented in Figure 2 for both laboratory X-rays and neutrons. Although the samples are mostly phase pure, some starting reagent Li$_{2}$CO$_{3}$ was found as an impurity in the neutron data, which is more of a bulk technique than X-ray diffraction. Furthermore, two very broad peaks could be indexed as close to the lattice parameters of the parent phase $\beta$-FeSe.  Indeed, these peaks have also been observed in previous work.$^{17, 19}$ The much broader peak width for the FeSe impurity is indicative of very small crystallite size and quantitatively fitting this phase is not possible given its nearly amorphous nature.

\begin{table}[b!]
\caption{Rietveld refinement of synchrotron PXRD data collected at 7 K for a superconducting sample of (${^{7}}$Li${_{1-x}}$Fe$_{x}$OD)FeSe shown in Figure 3 and a non-superconducting sample. Both samples are fitted to a $P4/nmm$ space group with origin choice 1. The tetrahedral angles $\alpha$$_{1}$ and $\alpha$$_{2}$ represent the Se-Fe-Se angles in and out of the basal plane, respectively.}
\resizebox{\columnwidth}{!}{
\begin{tabular}{l l l l l l l }
\hline
\hline
\multicolumn{7}{r} {$a = 3.7725$(1) \AA, $c = 9.1330$(2) \AA}, R$_{wp}$ = 12.83\%, $T_{c}$ = 37 K \\
\hline
\textbf{Atom}		&      \textbf{Wyckoff}  		& 	\textbf{\textit{x}}	&  \textbf{\textit{y}}		&	\textbf{\textit{z}}	&    \textbf{Occ.}		& \textbf{$U_{iso}$ (\AA$^2$)}	\\
\textbf{}		&      \textbf{site}  		& 	\textbf{\textit{}}	&  \textbf{\textit{}}		&	\textbf{\textit{}}		&	\textbf{\textit{}}		& \textbf{}	\\
\hline
Li/Fe1	&      2b		&	0  		&  	0 	 	&	0.5  			& 0.827/0.173(2)      &	0.0134				\\
Fe2		&      2a		&	0.5  		&  	0.5 	 	&	0 			& 0.979(2)               &	0.0057				\\
O		&      2c		&	0.5  		&  	0  	 	&	0.4266(3) 		& 1                         &	0.0037(7)				\\
Se		&      2c		&	0   		&  	0.5 	 	&	0.1603(1)		& 1                         &	0.0028(2)				\\
\hline
$\alpha$$_{1}$ ($^\circ$)	&      $\alpha$$_{2}$ ($^\circ$)			&	Fe-Fe (\AA)  		&  	Fe-Se (\AA) \\
104.38(2) 	&      112.07(1)	& 2.6675(1)	& 2.3875(4) 
\\
\hline
\hline
\multicolumn{7}{r} {$a = 3.7820$(1) \AA, $c = 9.0992$(1) \AA}, R$_{wp}$ = 10.66\%, non-superconducting \\
\hline
\textbf{Atom}		&      \textbf{Wyckoff}  		& 	\textbf{\textit{x}}	&  \textbf{\textit{y}}		&	\textbf{\textit{z}}	&    \textbf{Occ.}		& \textbf{$U_{iso}$ (\AA$^2$)}	\\
\textbf{}		&      \textbf{site}  		& 	\textbf{\textit{}}	&  \textbf{\textit{}}		&	\textbf{\textit{}}		&	\textbf{\textit{}}		& \textbf{}	\\
\hline
Li/Fe1	&      2b		&	0  		&  	0 	 	&	0.5  			& 0.809/0.191(2)      &	0.0156				\\
Fe2		&      2a		&	0.5  		&  	0.5 	 	&	0 			& 0.919(2)               &	0.0036				\\
O		&      2c		&	0.5  		&  	0  	 	&	0.4252(3) 		& 1                         &	0.0038(1)				\\
Se		&      2c		&	0   		&  	0.5 	 	&	0.1609(1)		& 1                         &	0.0019(6)				\\
\hline
$\alpha$$_{1}$ ($^\circ$)	&      $\alpha$$_{2}$ ($^\circ$)			&	Fe-Fe (\AA)  		&  	Fe-Se (\AA) \\
104.51(2) 	&      112.01(1)	& 2.6743(1)	& 2.3914(3) 
\\
\hline
\end{tabular}}
\label{crystal_data}
\end{table}

%%%%%%%%%%%%%%%%%%%%%%%%          Figure 3   %%%%%%%%%%%%%%%%%%%%%%%%%%%%%%%%%%%%%%%%%%
\begin{figure}[b]
\centering
  \includegraphics[width=1.05\columnwidth]{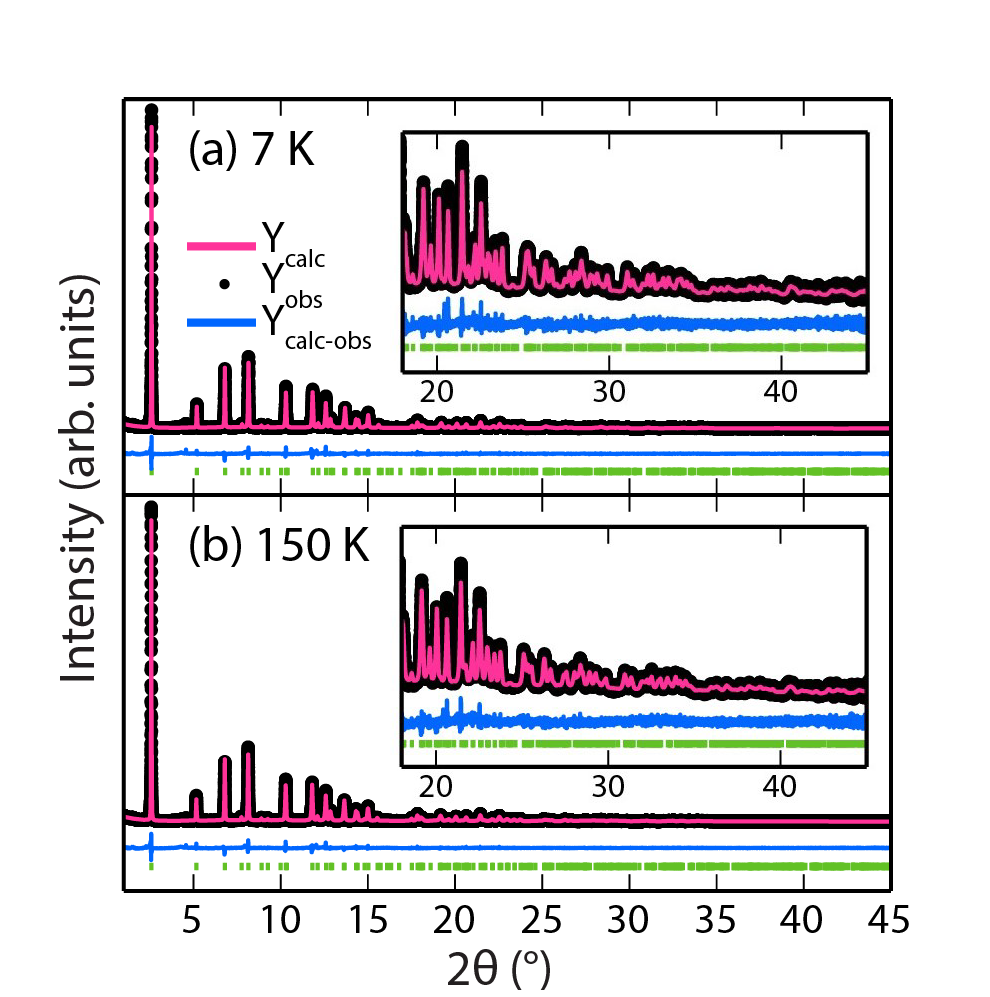}
  \caption{Synchrotron powder X-ray diffraction pattern at (a) 7 K and (b) 150 K for a single crystal sample of (Li${_{1-x}}$Fe$_{x}$OD)FeSe prepared at 120 $^\circ$C for 5 days ($T_{c}$ = 37 K ). Rietveld refinement of data collected at both temperatures did not reveal any lowering of symmetry from tetragonal $P4/nmm$. Tick marks representing the tetragonal phase are shown below the calculated, observed, and differences curves. The insets shown are a zoom in of the high-angle synchrotron data.}
  \label{fgr_3}
\end{figure}

The temperature-dependent synchrotron diffraction data did not reveal any major crystallographic changes in the structure (Figure 3). Therefore, unlike the parent FeSe phase, which undergoes a tetragonal to orthorhombic phase transition near 75 K,$^{29}$ (Li${_{1-x}}$Fe$_{x}$OH)FeSe and (Li${_{1-x}}$Fe$_{x}$OD)FeSe remain tetragonal down to base temperature (10 K).  Rietveld refinements of one of the deuteroxide patterns at 7 K and 150 K are presented in Figure 3, and relevant structural parameters are in Table 1 for both superconducting and non-superconducting deuteroxide phases.  Relevant bond distances and bond angles are also shown in Table 1. Only results from the synchrotron X-ray dataset are presented in Table 1, and structural parameters from the Rietveld refinements, including the neutron data, for the rest of the samples used to construct the full phase diagrams can be found in the ESI (Tables S1-S5).

%%%%%%%%%%%%%%%%%%%%%%%%%%%%%%%%%%%%%%%%%%%%%%%%%%%%%%%%%
\subsection{Magnetization results and the phase diagrams}

The SQUID magnetic susceptibility measurements for the series of hydroxide samples prepared through the powder routes are presented in Figure 4a. The deuteroxide samples, which were all derived from the single crystal route, are shown in Figure 4b. Only one sample within the hydroxide series expressed a ferromagnetic signal within the superconducting regime. A similar plot (Figure S1) for the hydroxide system prepared via the single crystal route can be found in the ESI.

%%%%%%%%%%%%%%%%%%%%%%%%          Figure4   %%%%%%%%%%%%%%%%%%%%%%%%%%%%%%%%%%%%%%%%%%
\begin{figure}[b]
\centering
  \includegraphics[width=1.0\columnwidth]{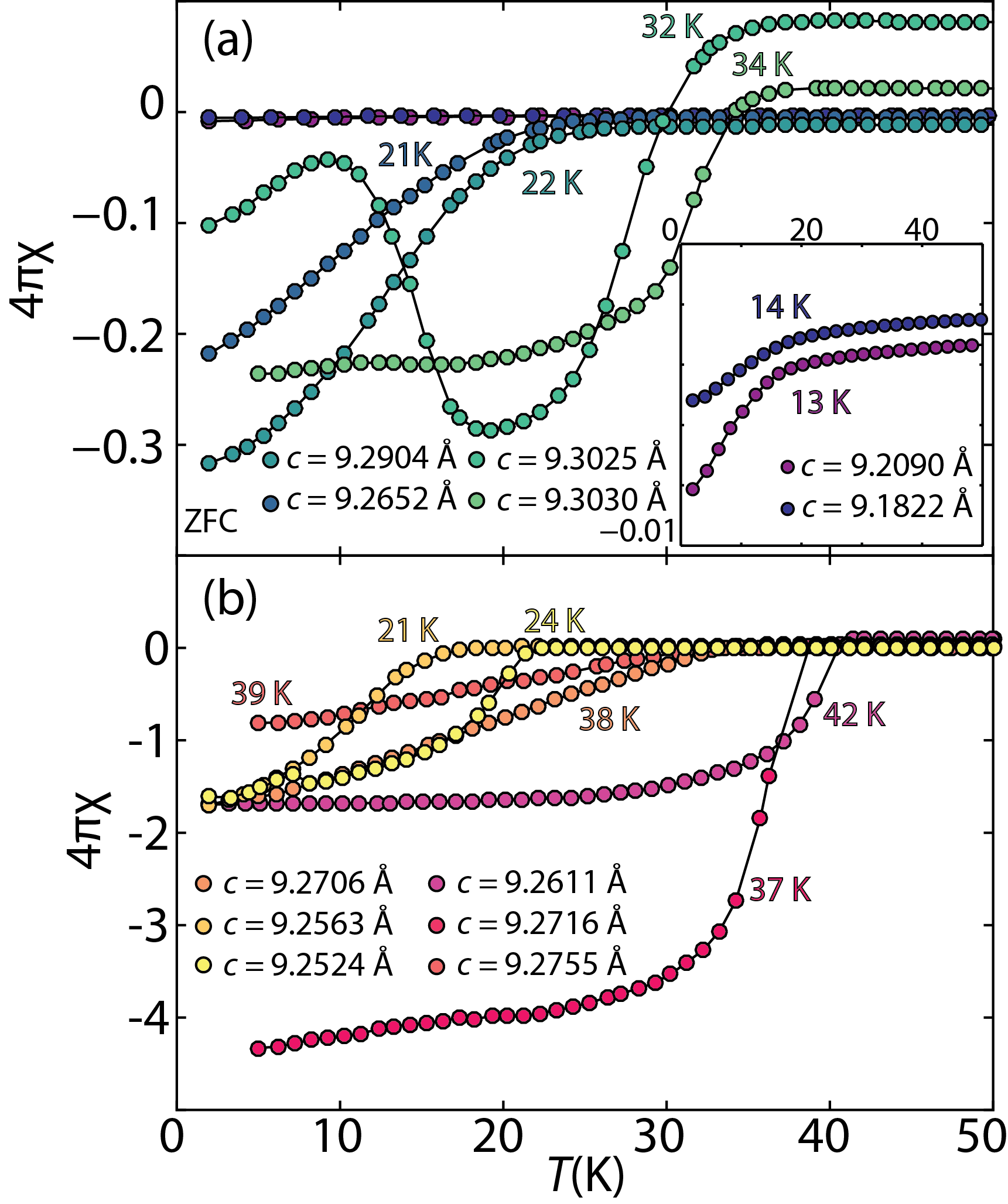}
  \caption{Magnetic susceptibility of samples prepared by (a) powder routes and (b) single crystal routes of deuteroxide series (Li${_{1-x}}$Fe$_{x}$OD)FeSe. For $T_{c}$ = 32 K, a ferromagnetic transition can be noted at $T_{f}$  = 10 K. ZFC data of powder samples and single crystal samples collected with applied fields of 1 mT and 3 mT, respectively, are shown.}
  \label{fgr_4}
\end{figure}

We have constructed superconducting phase diagrams in Figure 5 that relate the critical temperatures $T_{c}$ to the lattice constants compiled from the SQUID data and diffraction results from all the samples. The lattice parameters of the tetragonal unit cell found at room temperature were used as the x-axis versus $T_{c}$ in the phase diagrams. More specifically, we found the best correlation to $T_{c}$ is that of the tetragonality parameter, which is the simple $c/a$ ratio. The corresponding superconducting volume fractions (4$\pi$$\chi$) were also established by SQUID magnetometry (Figure 4). 
We found that $T_{c}$ and its volume fraction increased with the lattice constant $c$ and decreased with lattice constant $a$.  Therefore, those with the highest tetragonality gave the maximum $T_{c}$ and superconducting volume fractions. For samples to exhibit significant superconducting volume fractions (4$\pi$$\chi$ $>$ 10 \%), the lattice constant $c$ must be larger than about 9.20 \AA \ and $a$ smaller than about 3.80 \AA.  These trends in the lattice parameters are consistent with the findings of Sun $et$ $al$. on their hydroxide analogues,$^{18}$ where they attribute a large a lattice constant to iron vacancies in the FeSe layers and therefore iron slightly oxidized above 2+.

%%%%%%%%%%%%%%%%%%%%%%%%          Figure 5   %%%%%%%%%%%%%%%%%%%%%%%%%%%%%%%%%%%%%%%%%%
\begin{figure}[b]
\centering
  \includegraphics[width=1\columnwidth]{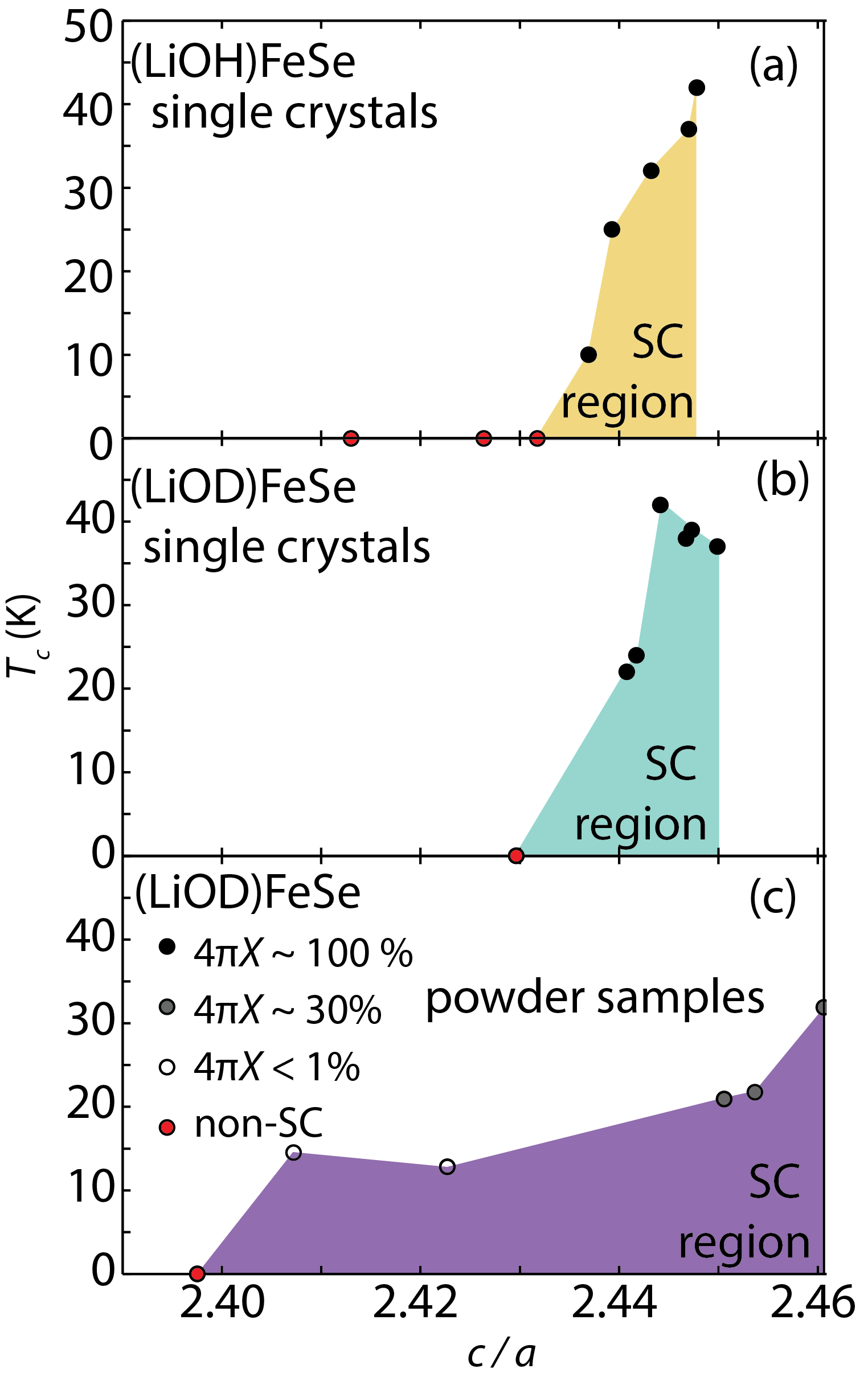}
  \caption{Superconducting phase diagrams of (a) (Li${_{1-x}}$Fe$_{x}$OH)FeSe  as comparison for that of (b) single crystal (${^{7}}$Li${_{1-x}}$Fe$_{x}$OD)FeSe and (c) powder samples and (Li${_{1-x}}$Fe$_{x}$OD)FeSe.  The critical temperatures Tc are related to the tetragonality parameter, which is the simple ratio of the lattice parameters $c/a$.}
  \label{fgr_5}
\end{figure}

Our combined diffraction experiments did indeed find variations on the iron occupancies, whether superconducting or non-superconducting. In general, the smaller the tetragonality parameter, the lower the $T_{c}$.  As Table 1 shows, when the occupancy of the Fe2 site falls from near full to 91.9(2) \%, superconductivity is lost. The differences overall between the hydroxide samples prepared by powder routes and the single crystal ones could be due the accommodation of iron vacancies during the syntheses. As Sun $e$t $al$. found in their samples,$^{28}$ when FeSe in its tetragonal $\beta$-phase is used as the host for intercalation via hydrothermal synthesis, iron powder is necessary in order to fill in the resulting vacancies.  When we start with Fe powder and selenourea as the Se source, this leads to more variability in the amount of iron vacancies and therefore a larger spread in the tetragonality parameter that can express superconductivity (Figure 5). Our powder method therefore would lead to the in-situ growth of alternating FeSe and Li${_{1-x}}$Fe${_{x}}$OD layers rather than post-synthetic modification (also known as soft chemistry) of FeSe layers as done in our single crystal method.

%%%%%%%%%%%%%%%%%%%%%%%%%%%%%%%%%%%%%%%%%%%%%%%%
\subsection{Neutron results}

To verify whether any of the samples exhibit antiferromagnetism, we searched for any superlattice peaks in the NPD patterns that could arise below 120 K, the AFM transition in the parent phase of (Li${_{1-x}}$Fe${_{x}}$OD)FeSe reported by Dong $et$ $al$.$^{1}$ No superlattice reflections were observed in the BT-1 high-resolution NPD patterns, and our deuterated samples allowed for a low background in case of a small Fe signal. Indeed, in the arsenide systems the iron moment can be small in the parent phases such as 0.36(5) $\mu$$_{B}$/Fe in LaOFeAs30 and 0.25(7) $\mu$$_{B}$/Fe in NdOFeAs,$^{31}$ Any hydrogen incoherent background would easily overwhelm such a small signal from long-range magnetic ordering in the NPD.
The samples were measured up to 50 K on BT-1, and no long range magnetic ordering was observed. NPD patterns measured with a PSD on BT-7, which has a much higher flux than BT-1 at low angles, also revealed no antiferromagnetic peaks in the non-superconducting samples (Figure 6).  Difference patterns between 150 K and 4 K are shown in Figure 6, revealing no residual intensity and only differences arising from thermal expansion and thus peak positions.

%%%%%%%%%%%%%%%%%%%%%%%%%%%%%%%%%%%%%%%%%%%%%%%%%%%%%%%%%

\section{Discussion}

\subsection{Relation between structural parameters and superconductivity}

In preparing our deuteroxide samples, we found the reaction temperature to influence the lattice constants. Mild hydrothermal reaction temperatures (120 $^\circ$C) led to samples expressing a higher $T_{c}$, while the reaction temperature above 180 $^\circ$C led exclusively to either non-superconducting samples or ones with very low volume fractions (4$\pi$$\chi$$<$1\%). Reaction times also affected the lattice constants. Longer reaction times ($>$3 days) yielded samples with slightly larger $a$ and smaller $c$ (i.e. smaller tetragonality parameters).
While all the deuterated samples followed the trend shown in the phase diagram (Figure 5), similarly prepared hydrated samples deviated in their behaviour.  Indeed, some hydroxide samples with lattice parameters matching those in the phase diagram from earlier literature1 did not exhibit superconductivity (Figure 4a).
Interestingly, samples prepared at lower temperatures with the described mixing ratio and longer reaction times expressed coexistence between superconductivity and ferromagnetism (Figure 3). Thus, while longer reaction times above 180 $^\circ$C led to lower $T_{c}$'s or non-superconducting samples, longer reaction times at lower temperatures (120 $^\circ$C) produced the ferromagnetic signal in superconducting (Li${_{1-x}}$Fe${_{x}}$OD)FeSe.  As described in the next section, this might be a kinetic effect from the increasing amount of oxidized iron in water from longer reaction times.

%%%%%%%%%%%%%%%%%%%%%%%%%%%%%%%%%%%%%%%%%%%%%%%%%%%%%%%%%
\subsection{Relation between magnetism and superconductivity}

In several of our non-superconducting samples, we have observed an antiferromagnetic transition close to 120 K. Dong et al. have claimed that the hydroxide samples in the non-superconducting dome were antiferromagnetic parent phases with a $T_{N}$ close to 120 K, and therefore that the selenides and arsenides have the same underlying physics with respect to the superconducting mechanism.  This is a very important claim that could have large implications in the field of iron-based superconductors. None of our non-superconducting deuteroxide samples, however, exhibited this antiferromagnetic signal in the SQUID measurements, which led us to believe that the antiferromagnetism may not be intrinsic to this system.

%%%%%%%%%%%%%%%%%%%%%%%%          Figure 6   %%%%%%%%%%%%%%%%%%%%%%%%%%%%%%%%%%%%%%%%%%
\begin{figure}[b]
\centering
  \includegraphics[width=1\columnwidth]{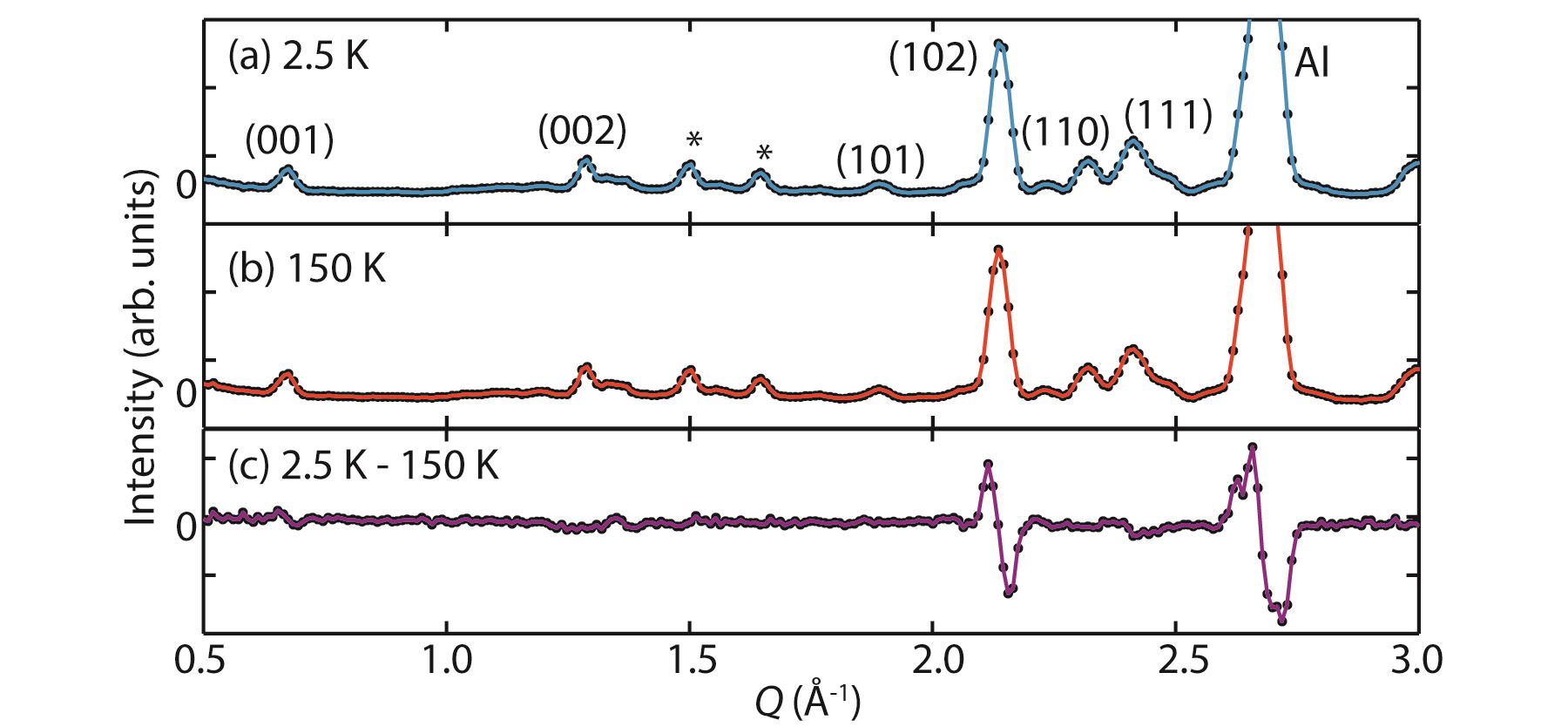}
  \caption{(a) The NPD pattern of non-superconducting phase of (${^{7}}$Li${_{1-x}}$FexOD)FeSe at 150 K and (b) 4 K.  The difference between the two patterns in (c) reveals no antiferromagnetic peaks.}
  \label{fgr_6}
\end{figure}

Our findings in the preparation of hydroxide and deuteroxide samples revealed the strong possibility that the 120 K transition observed in the SQUID magnetization measurements arise from iron oxide impurities. The so-called Verwey transition, which corresponds to a Fe$^{2+}$/Fe$^{3+}$  charge ordering transition in Fe${_{3}}$O${_{4}}$ also occurs near 120 K.$^{32, 33}$ Furthermore, structurally related Li${_{x}}$FeO${_{2}}$ phases can express $T_{N}$ from 100 K to 300 K according to amount of intercalated lithium cations.$^{34, 35}$  

In order to study the formation of iron oxide impurities, a sample was prepared under similar hydrothermal conditions but without the addition of selenourea. As pointed out by Sun $et$ $al$. in their extensive study of the formation of (Li${_{1-x}}$Fe${_{x}}$OH)FeSe, the strongly basic conditions (pH $>$ 14) of the synthesis strongly favors the formation of Fe$^{3+}$ species according the electrochemical-pH phase equilibrium diagram (i.e. Pourbaix) of iron.$^{36}$ Therefore, without the selenourea reagant to stabilize divalent iron, a large amount of mixed valent iron oxides are produced from hydrothermal synthesis containing large amounts of LiOH (or LiOD).

%%%%%%%%%%%%%%%%%%%%%%%%          Figure 7   %%%%%%%%%%%%%%%%%%%%%%%%%%%%%%%%%%%%%%%%%%
\begin{figure}[b]
\centering
  \includegraphics[width=1\columnwidth]{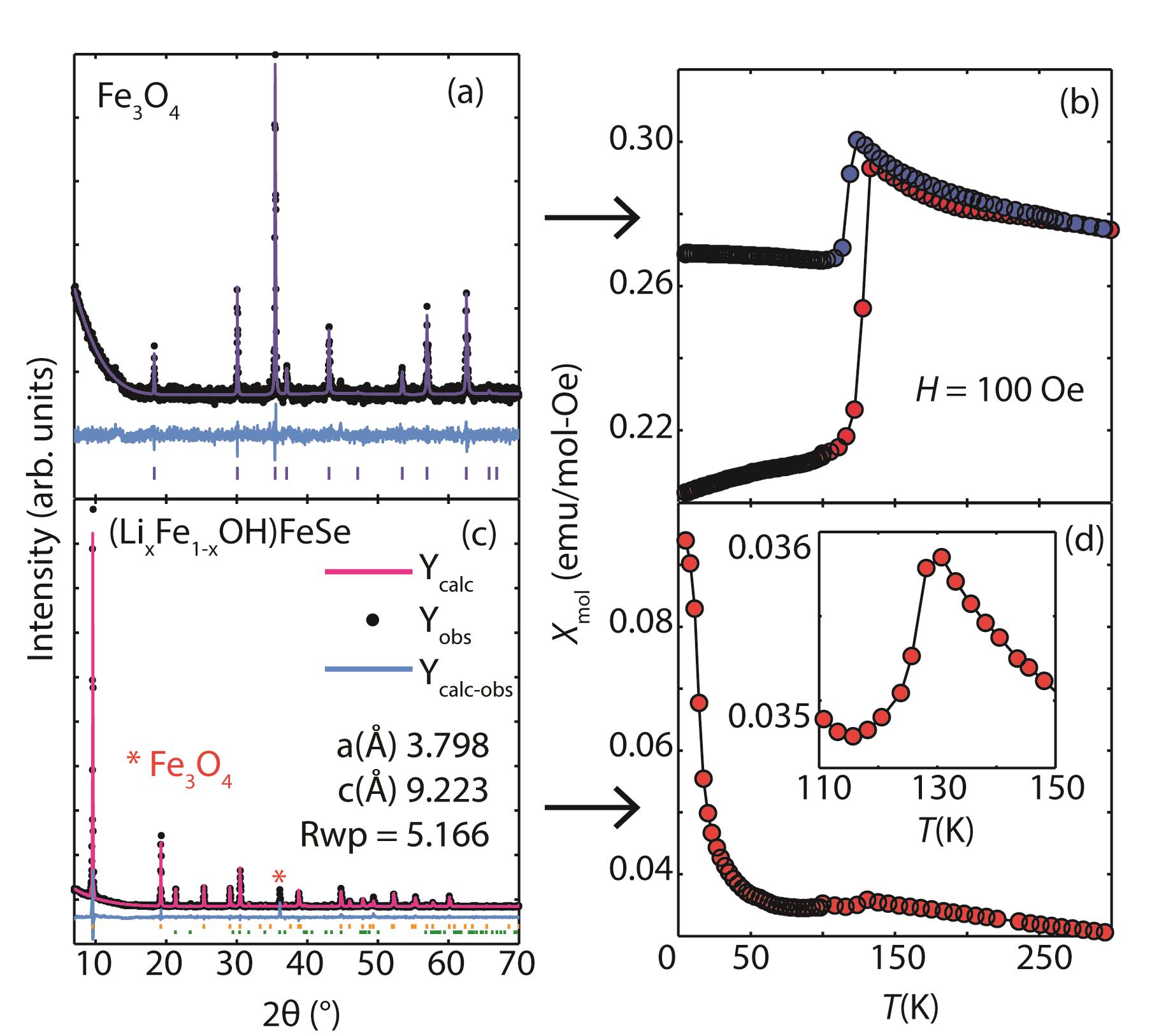}
  \caption{(a) Powder XRD and Rietveld analysis of phase pure Fe${_{3}}$O${_{4}}$ prepared under similar hydrothermal conditions to that of (Li${_{1-x}}$Fe$_{x}$OH)FeSe in the absence of selenourea. (b) The corresponding magnetization data of the Fe${_{3}}$O${_{4}}$ sample indicating the Verwey transition near 125 K. (c) The powder XRD of (Li${_{1-x}}$Fe$_{x}$OH)FeSe with the Fe${_{3}}$O${_{4}}$ impurity marked along its strongest reflection.  (d) The corresponding magnetization data of this non-superconducting (Li${_{1-x}}$Fe$_{x}$OH)FeSe sample, exhibiting the 125 K transition similar to that of Fe$^{2+}$/Fe$^{3+}$ charge ordering seen in Fe${_{3}}$O${_{4}}$ (inset).}
  \label{fgr_6}
\end{figure}

The XRD pattern of the as-recovered sample from hydrothermal synthesis without selenourea was fitted to the Fe${_{3}}$O${_{4}}$ structure, and its magnetic susceptibility measurement was in very good agreement with typical Verwey transition at 120 K (Figure 7). As shown in Figure 7b, a hydroxide sample with lattice constants in the supposed superconducting region showed no superconductivity, but a transition similar to charge ordering in Fe${_{3}}$O${_{4}}$. Peaks in the XRD pattern (indicated by * in Figure 7b) that cannot be matched with the (Li${_{1-x}}$Fe${_{x}}$OH)FeSe phase was indexed well with the strongest peaks of Fe${_{3}}$O${_{4}}$. In addition, our synchrotron XRD data for a non-superconducting (Li${_{1-x}}$Fe${_{x}}$OH)FeSe single crystal sample revealed small amounts of Fe${_{3}}$O${_{4}}$ impurity (Figure S2), which was not observed by laboratory X-ray measurements. Therefore, it is likely that the 120 K transition in non-superconducting (Li${_{1-x}}$Fe${_{x}}$OH)FeSe samples is extrinsic and due to magnetic impurities Fe${_{3}}$O${_{4}}$ or structurally related Li${_{x}}$FeO${_{2}}$, which would modulate the ordering temperature.  

Hydrothermal synthesis of samples with D${_{2}}$O under similar conditions as those with H${_{2}}$O did not lead to appreciable oxide impurity.  We therefore conclude that the observed differences in the acid-base chemistry of H${_{2}}$O and D${_{2}}$O lead to different products for similar reaction conditions. Indeed, a hydrothermal treatment of iron powder with D${_{2}}$O and with LiOD did not lead to complete conversion to Fe${_{3}}$O${_{4}}$ but left unreacted iron powder (approximately 50\%). Since under highly basic conditions, Fe${^{3+}}$ should be favoured thermodynamically, we believe the kinetics for the oxidation of iron with D${_{2}}$O is slower than in H${_{2}}$O. The autoionization constant of D${_{2}}$O is smaller than that of H${_{2}}$O due to the stronger D–O bond than the H–O bond. 

All deuteroxide single crystal samples prepared in D${_{2}}$O showed noticeable higher $T_{c}$'s than their hydroxide counterparts prepared at identical conditions (Table S2). The $T_{c}$ of hydroxide samples can be improved by reducing the reaction time (i.e. 37 K vs. 32 K for 2 d and 4 d, respectively at 120 $^\circ$C). It is likely that a shorter reaction time reduced the extent of Fe oxidation. Interestingly, both (Li${_{1-x}}$Fe${_{x}}$OH)FeSe and (Li${_{1-x}}$Fe${_{x}}$OD)FeSe single crystals prepared using Sn metal instead of Fe powder at 120 $^\circ$C showed the same $T_{c}$ at 42 K, higher than other samples without using Sn. The advantage of Sn metal was to create a reducing environment without introducing iron oxide impurities, due to lack of Fe powder.

Although we have established here that the (Li${_{1-x}}$Fe${_{x}}$OH)FeSe system likely does not have a parent antiferromagnetic phase, we do not suggest that the chalcogenide-based
superconductors are not linked to the arsenide-based superconductors from the present results.  The Fe${_{1+x}}$Te${_{y}}$Ch${_{1-y}}$ for $Ch$ = Se and S phases in particular exhibit a rich magnetic phase diagram$^{37,38}$ before superconductivity sets in with chemical substitution.$^{39}$ The ordered vacancy phase of K${_{0.8}}$Fe${_{1.6}}$Se${_2}$.$^{24, 25}$ has also shown an antiferromagnetic transition at large temperatures (about 559 K)$^{40}$ while the disordered vacancy phase exhibits a $T_{c}$ close to 30 K.$^{41-45}$  What distinguishes those two systems, however, from the present compound and FeSe, is the lack of a large magnetic moment on iron in the latter compounds.$^{46-51}$ In Fe${_{1+x}}$Te it can be as large as 2 $\mu$$_{B}$$^{52}$ and in K${_{0.8}}$Fe${_{1.6}}$Se${_2}$ as large as 3.3(1) $\mu$$_{B}$.$^{40}$ Not surprisingly, when in the superconducting regime, both compounds exhibit a spin resonance energy in the inelastic neutron spectra, which corresponds to spin fluctuations.  A large magnetic moment is clearly not the case in the present system.

As to the ferromagnetic transition observed at about 10 K in the sample with a $T_{c}$ of 34 K, several authors have also observed it in in the hydroxide analogues.  Pachmeyer $et$ $al$. attribute the long range magnetic order at 18 K to the iron cations partially substituted on the Li site,$^{17}$ while Lu $et$ $al$. assign this transition as being antiferromagnetic (about 12 K) according to their NMR studies.$^{19}$ Our recent small angle neutron scattering study illustrates the formation of long-range magnetic order below 12.5 K, but with a moment too small to see with diffraction.$^{22}$ No doubt this observation arises from the crystallographic site where the moment is located is too dilute with iron occupancy.

\section{Conclusions}
In conclusion, we have successfully mapped out a phase diagram for (Li${_{1-x}}$Fe${_{x}}$OD)FeSe and have found that the highest $T_{c}$ for deuterated samples is 42 K, and the $T_{c}$ for both deuterated and hydroxide samples correlate with lattice constants. Since the highest $T_{c}$ observed for the hydroxide sample was also approximately 42 K, we conclude that there is no isotope effect on the superconducting properties in substituting H by D.  Mild hydrothermal preparation for long reaction times can lead to the coexistence of ferromagnetism and superconductivity. Finally, any claims of anti-ferromagnetism in the parent phase of this system should be re-evaluated in light of the easy preparation of oxide impurities with transition temperatures near the vicinity of 120 K in H${_{2}}$O.

\section{Acknowledgements}
Research at the University of Maryland was supported by the NSF Career DMR-1455118 and the AFOSR Grant No. FA9550-14-10332. We acknowledge the support of the National Institute of Standards and Technology, U. S. Department of Commerce, in providing the neutron research facilities used in this work.  Use of the Advanced Photon Source at Argonne National Laboratory was supported by the U. S. Department of Energy, Office of Science, Office of Basic Energy Sciences, under Contract No. DE-AC02-06CH11357.  We thank S. Lapidus for his assistance with the 11-BM measurements.

%%%END OF MAIN TEXT%%%

\textbf{}

\textbf{}

\textbf{REFERENCES}

\textbf{}

1.	X. Dong, H. Zhou, H. Yang, J. Yuan, K. Jin, F. Zhou, D. Yuan, L. Wei, J. Li and X. Wang, \textit{J. Am. Chem. Soc.}, 2014, \textbf{137}, 66-69.

2.	J. Paglione and R. L. Greene, \textit{Nat. Phys.}, 2010, \textbf{6}, 645-658.

3.	D. C. Johnston, \textit{Adv. Phys.}, 2010, \textbf{59}, 803-1061.

4.	T. M. McQueen, Q. Huang, V. Ksenofontov, C. Felser, Q. Xu, H. Zandbergen, Y. S. Hor, J. Allred, A. J. Williams, D. Qu, J. Checkelsky, N. P. Ong and R. J. Cava, \textit{Phys. Rev. B}, 2009, \textbf{79}, 014522.

5.	F.-C. Hsu, J.-Y. Luo, K.-W. Yeh, T.-K. Chen, T.-W. Huang, P. M. Wu, Y.-C. Lee, Y.-L. Huang, Y.-Y. Chu, D.-C. Yan and M.-K. Wu, \textit{Proc. Natl. Acad. Sci. U.S.A.}, 2008, \textbf{105}, 14262-14264.

6.	H. Kotegawa, S. Masaki, Y. Awai, H. Tou, Y. Mizuguchi and Y. Takano, \textit{J. Phys. Soc. Jpn.}, 2008, \textbf{77}, 113703.

7.	S. Margadonna, Y. Takabayashi, M. T. McDonald, K. Kasperkiewicz, Y. Mizuguchi, Y. Takano, A. N. Fitch, E. Suard and K. Prassides, \textit{Chem. Comm.}, 2008, 5607-5609.

8.	S. Medvedev, T. McQueen, I. Troyan, T. Palasyuk, M. Eremets, R. Cava, S. Naghavi, F. Casper, V. Ksenofontov and G. Wortmann, \textit{Nat. Mater.}, 2009, \textbf{8}, 630-633.

9.	T. Imai, K. Ahilan, F. L. Ning, T. M. McQueen and R. J. Cava, \textit{Phys. Rev. Lett.}, 2009, \textbf{102}, 177005.

10.	E.-W. Scheidt, V. Hathwar, D. Schmitz, A. Dunbar, W. Scherer, F. Mayr, V. Tsurkan, J. Deisenhofer and A. Loidl, \textit{Eur. Phys. J. B}, 2012, \textbf{85}, 1-5.

11.	M. Burrard-Lucas, D. G. Free, S. J. Sedlmaier, J. D. Wright, S. J. Cassidy, Y. Hara, A. J. Corkett, T. Lancaster, P. J. Baker, S. J. Blundell and S. J. Clarke, \textit{Nat. Mater.}, 2013, \textbf{12}, 15-19.

12.	T. Ying, X. Chen, G. Wang, S. Jin, X. Lai, T. Zhou, H. Zhang, S. Shen and W. Wang, \textit{J. Am. Chem. Soc.}, 2013, \textbf{135}, 2951-2954.

13.	S. J. Sedlmaier, S. J. Cassidy, R. G. Morris, M. Drakopoulos, C. Reinhard, S. J. Moorhouse, D. O’Hare, P. Manuel, D. Khalyavin and S. J. Clarke, \textit{J. Am. Chem. Soc.}, 2014, \textbf{136}, 630-633.

14.	J.-F. Ge, Z.-L. Liu, C. Liu, C.-L. Gao, D. Qian, Q.-K. Xue, Y. Liu and J.-F. Jia, \textit{Nat. Mater.}, 2015, \textbf{14}, 285-289.

15.	S. He, J. He, W. Zhang, L. Zhao, D. Liu, X. Liu, D. Mou, Y.-B. Ou, Q.-Y. Wang and Z. Li, \textit{Nat. Mater.}, 2013, \textbf{12}, 605-610.

16.	S. Tan, Y. Zhang, M. Xia, Z. Ye, F. Chen, X. Xie, R. Peng, D. Xu, Q. Fan and H. Xu, \textit{Nat. Mater.}, 2013, \textbf{12}, 634-640.

17.	U. Pachmayr, F. Nitsche, H. Luetkens, S. Kamusella, F. Brückner, R. Sarkar, H.-H. Klauss and D. Johrendt, \textit{Angew. Chem. Int. Ed.}, 2015, \textbf{54}, 293-297.

18.	H. Sun, D. N. Woodruff, S. J. Cassidy, G. M. Allcroft, S. J. Sedlmaier, A. L. Thompson, P. A. Bingham, S. D. Forder, S. Cartenet, N. Mary, S. Ramos, F. R. Foronda, B. H. Williams, X. Li, S. J. Blundell and S. J. Clarke, \textit{Inorg. Chem.}, 2015, \textbf{54}, 1958-1964.

19.	X. F. Lu, N. Z. Wang, H. Wu, Y. P. Wu, D. Zhao, X. Z. Zeng, X. G. Luo, T. Wu, W. Bao, G. H. Zhang, F. Q. Huang, Q. Z. Huang and X. H. Chen, \textit{Nat. Mater.}, 2015, \textbf{14}, 325-329.

20.	X. F. Lu, N. Z. Wang, G. H. Zhang, X. G. Luo, Z. M. Ma, B. Lei, F. Q. Huang and X. H. Chen, \textit{Phys. Rev. B}, 2014, \textbf{89}, 020507.

21.	U. Pachmayr and D. Johrendt, \textit{Chem. Comm.}, 2015, \textbf{51}, 4689-4692.

22.	J. W. Lynn, X. Zhou, C. K. H. Borg, S. R. Saha, J. Paglione and E. E. Rodriguez, \textit{Phys. Rev. B}, 2015, \textbf{92}, 060510.

23.	X. Dong, K. Jin, D. Yuan, H. Zhou, J. Yuan, Y. Huang, W. Hua, J. Sun, P. Zheng and W. Hu, \textit{Phys. Rev. B}, 2015, \textbf{92}, 064515.

24.	Y. Mizuguchi, H. Takeya, Y. Kawasaki, T. Ozaki, S. Tsuda, T. Yamaguchi and Y. Takano, \textit{Appl. Phys. Lett.}, 2010, \textbf{98}, 042511.

25.	J. Ying, X. Wang, X. Luo, A. Wang, M. Zhang, Y. Yan, Z. Xiang, R. Liu, P. Cheng and G. Ye, \textit{Phys. Rev. B}, 2011, \textbf{83}, 212502.

26.	P. G. Harrison, \textit{Chemistry of tin, Blackie}, 1989.

27.	J. Lynn, Y. Chen, S. Chang, Y. Zhao, S. Chi, W. Ratcliff, B. Ueland and R. Erwin, \textit{J. Res. NIST}, 2012, \textbf{117}, 61-79.

28.	R. W. Cheary and A. Coelho, \textit{J. Appl. Crystallogr.}, 1992, \textbf{25}, 109.

29.	T. McQueen, A. Williams, P. Stephens, J. Tao, Y. Zhu, V. Ksenofontov, F. Casper, C. Felser and R. Cava, \textit{Phys. Rev. Lett.}, 2009, \textbf{103}, 057002.

30.	C. de La Cruz, Q. Huang, J. Lynn, J. Li, W. Ratcliff II, J. L. Zarestky, H. Mook, G. Chen, J. Luo and N. Wang, \textit{Nature}, 2008, \textbf{453}, 899-902.

31.	Y. Chen, J. Lynn, J. Li, G. Li, G. Chen, J. Luo, N. Wang, P. Dai, C. dela Cruz and H. Mook, \textit{Phys. Rev. B}, 2008, \textbf{78}, 064515.

32.	F. Walz, \textit{J Phys.: Condens. Matter}, 2002, \textbf{14}, R285.

33.	Z. Zhang and S. Satpathy, \textit{Phys. Rev. B}, 1991, \textbf{44}, 13319.

34.	M. Tabuchi, K. Ado, H. Kobayashi, I. Matsubara, H. Kageyama, M. Wakita, S. Tsutsui, S. Nasu, Y. Takeda and C. Masquelier, \textit{J. Solid State Chem.}, 1998, \textbf{141}, 554-561.

35.	M. Tabuchi, S. Tsutsui, C. Masquelier, R. Kanno, K. Ado, I. Matsubara, S. Nasu and H. Kageyama, \textit{J. Solid State Chem.}, 1998, \textbf{140}, 159-167.

36.	J.-M. R. Genin, G. Bourrie, F. Trolard, M. Abdelmoula, A. Jaffrezic, P. Refait, V. Maitre, B. Humbert and A. Herbillon, \textit{Environ. Sci. Technol.}, 1998, \textbf{32}, 1058-1068.

37.	E. E. Rodriguez, C. Stock, P. Zajdel, K. L. Krycka, C. F. Majkrzak, P. Zavalij and M. A. Green, \textit{Phys. Rev. B}, 2011, \textbf{84}, 064403.

38.	P. Zajdel, P.-Y. Hsieh, E. E. Rodriguez, N. P. Butch, J. D. Magill, J. Paglione, P. Zavalij, M. R. Suchomel and M. A. Green, \textit{J. Am. Chem. Soc.}, 2010, \textbf{132}, 13000-13007.

39.	Y. Mizuguchi and Y. Takano, \textit{J. Phys. Soc. Jpn.}, 2010, \textbf{79}, 102001.

40.	W. Bao, Q.-Z. Huang, G.-F. Chen, M. A. Green, D.-M. Wang, J.-B. He and Y.-M. Qiu, \textit{Chin. Phys. Lett.}, 2011, \textbf{28}, 086104.

41.	J. Guo, S. Jin, G. Wang, S. Wang, K. Zhu, T. Zhou, M. He and X. Chen, \textit{Phys. Rev. B}, 2010, \textbf{82}, 180520.

42.	W. Hang-Dong, D. Chi-Heng, L. Zu-Juan, M. Qian-Hui, Z. Sha-Sha, F. Chun-Mu, H. Q. Yuan and F. Ming-Hu, \textit{Eur. Phys. Lett.}, 2011, \textbf{93}, 47004.

43.	X. G. Luo, X. F. Wang, J. J. Ying, Y. J. Yan, Z. Y. Li, M. Zhang, A. F. Wang, P. Cheng, Z. J. Xiang, G. J. Ye, R. H. Liu and X. H. Chen, \textit{New J. Phys.}, 2011, \textbf{13}, 053011.

44.	F. Ming-Hu, W. Hang-Dong, D. Chi-Heng, L. Zu-Juan, F. Chun-Mu, C. Jian and H. Q. Yuan, \textit{Eur. Phys. Lett.}, 2011, \textbf{94}, 27009.

45.	A. F. Wang, J. J. Ying, Y. J. Yan, R. H. Liu, X. G. Luo, Z. Y. Li, X. F. Wang, M. Zhang, G. J. Ye, P. Cheng, Z. J. Xiang and X. H. Chen, \textit{Phys. Rev. B}, 2011, \textbf{83}, 060512.

46.	V. Y. Pomjakushin, D. Sheptyakov, E. Pomjakushina, A. Krzton-Maziopa, K. Conder, D. Chernyshov, V. Svitlyk and Z. Shermadini, \textit{Phys. Rev. B}, 2011, \textbf{83}, 144410.

47.	D. Wang, J. He, T.-L. Xia and G. Chen, \textit{Phys. Rev. B}, 2011, \textbf{83}, 132502.

48.	P. Zavalij, W. Bao, X. Wang, J. Ying, X. Chen, D. Wang, J. He, X. Wang, G. Chen and P.-Y. Hsieh, \textit{Phys. Rev. B}, 2011, \textbf{83}, 132509.

49.	D. P. Shoemaker, D. Y. Chung, H. Claus, M. C. Francisco, S. Avci, A. Llobet and M. G. Kanatzidis, \textit{Phys. Rev. B}, 2012, \textbf{86}, 184511.

50.	X. Ding, D. Fang, Z. Wang, H. Yang, J. Liu, Q. Deng, G. Ma, C. Meng, Y. Hu and H.-H. Wen, \textit{Nat. Commun.}, 2013, \textbf{4}, 1897.

51.	D. Guterding, H. O. Jeschke, P. Hirschfeld and R. Valenti, \textit{Phys. Rev. B}, 2015, \textbf{91}, 041112.

52.	W. Bao, Y. Qiu, Q. Huang, M. Green, P. Zajdel, M. Fitzsimmons, M. Zhernenkov, S. Chang, M. Fang and B. Qian, \textit{Phys. Rev. Lett.}, 2009, \textbf{102}, 247001.

\end{document}